\documentclass[conference]{IEEEtran}
\IEEEoverridecommandlockouts
\usepackage{cite}

\ifCLASSINFOpdf
   \usepackage[pdftex]{graphicx}
\else
\fi
\usepackage{amsmath}
\usepackage{array}

\usepackage{multirow}

\usepackage{url}

\usepackage{xcolor}

\newcommand{\THETAVEC}[1]{\ensuremath{\overrightarrow{\theta}_{#1}}}

\begin{document}
%
\title{Mode connectivity in the QCBM loss landscape: ICCAD Special Session Paper \thanks{This manuscript has been authored by UT-Battelle, LLC under Contract No. DE-AC05-00OR22725 with the U.S. Department of Energy. The United States Government retains and the publisher, by accepting the article for publication, acknowledges that the United States Government retains a non-exclusive, paid-up, irrevocable, worldwide license to publish or reproduce the published form of this manuscript, or allow others to do so, for United States Government purposes. The Department of Energy will provide public access to these results of federally sponsored research in accordance with the DOE Public Access Plan. (http://energy.gov/downloads/doe-public-279 access-plan).}}

\author{\IEEEauthorblockN{Kathleen E. Hamilton\IEEEauthorrefmark{1},
Emily Lynn\IEEEauthorrefmark{2},
Vicente Leyton-Ortega\IEEEauthorrefmark{1}, 
Swarnadeep Majumder\IEEEauthorrefmark{3} and
Raphael C. Pooser\IEEEauthorrefmark{1}}
\IEEEauthorblockA{\IEEEauthorrefmark{1}Computational Sciences and Engineering Division\\
Oak Ridge National Laboratory,
Oak Ridge, Tennessee 37831\\
Email:  hamiltonke@ornl.gov}
\IEEEauthorblockA{\IEEEauthorrefmark{2}Department of Physics and Astronomy\\
Taylor University, 
Upland, Indiana 46989}
\IEEEauthorblockA{\IEEEauthorrefmark{3}Department of Electrical and Computer Engineering\\
Duke University, 
Durham, North Carolina 27708}}




\maketitle

\begin{abstract}
Quantum circuit Born machines (QCBMs) and training via variational quantum algorithms (VQAs) are key applications for near-term quantum hardware. QCBM ans\"atze designs are unique in that they do not require prior knowledge of a physical Hamiltonian. Many ans\"atze are built from fixed designs.  In this work, we train and compare the performance of QCBM models built using two commonly employed parameterizations and two commonly employed entangling layer designs.  In addition to comparing the overall performance of these models, we look at features and characteristics of the loss landscape --connectivity of minima in particular -- to help understand the advantages and disadvantages of each design choice. We show that the rotational gate choices can improve loss landscape connectivity.

\end{abstract}


%
\IEEEpeerreviewmaketitle


\section{Introduction}
\label{sec:introduction}
Parameterized quantum circuits (PQCs), which are used in both quantum circuit Born machines (QCBMs) \cite{benedetti2018generative,liu2018differentiable} and variational quantum algorithms (VQAs)\cite{VQA2021}, underpin the majority of applications that are executed on noisy intermediate-scale quantum (NISQ) hardware today. 
In both cases, circuit parameters are adjusted via a training algorithm which often seeks to minimize a cost function using data provided via quantum measurements. In many VQAs, the cost function and ansatz are often based on a Hamiltonian of a quantum system being simulated \cite{Abrams1999}. In the case of QCBMs, the ansatz can be abstract in that it need not express a wavefunction corresponding to a particular Hamiltonian, but rather is chosen to ensure sufficient expressiveness to represent a target distribution. The distribution is semantically encoded into the quantum state of the quantum register, and measurement of the quantum register essentially samples from it.\cite{hamilton2019generative,leyton2019robust,Benedetti2019ddqcl, Benedetti2019PQC}.
We study the trainability of several parameterizations and entangling layout designs by examining the loss landscape under various conditions. We leverage visualization techniques commonly employed in classical machine learning ~\cite{goodfellow2014qualitatively,li2017visualizing,Hamilton2020visualization} to study the loss landscape in lower dimensions in order to identify minima and various features, such as ravines, and extrapolate the inter-connectedness of these features in higher dimensions. We use the nudged elastic band \cite{henkelman2000neb}  to uncover the low loss paths between minima features to determine if they are interconnected. 

The main contribution of our work is a quantitative study of the trade-off between entangling layer sparseness, parameterization, and circuit depth in terms of their impact on trainability.  We highlight several design aspects for QCBMs that should be considered as these circuits are scaled up in width (number of qubits increases), and depth (number of entangling layers and trainable parameters increases).  Wider circuits with sparse entangling layers lose sufficient expressiveness to appreciably train, while sparse parameterizations lead to flatter, less interconnected landscapes and high variance in the converged distributions.  Our study relies on brute force exploration of the landscape with 2-, 4- and 6-qubit models and we present proof of concept results that show that large barriers can exist in the loss landscapes, and that there may be low loss paths around these barriers.

\section{Methods}
\label{sec:methods}
\subsection{QCBM construction}
\label{sec:qcbm_construction}
QCBMs are quantum circuit models that have been used in variational algorithms, where the objective is to prepare a target distribution with high fidelity.  A quantum state is prepared by a parameterized ansatz $\mathcal{U}(\THETAVEC{})$ which takes the initial qubit register from the all zero state $| \psi_0 \rangle = |000\dots 0\rangle$ to a final state $|\psi(\THETAVEC{})\rangle = \mathcal{U}(\THETAVEC{})|\psi_0\rangle$.  The final state is sampled in the computational basis, and the final result in a classical probability distribution $Q(\THETAVEC{})$ defined over the $2^n$ computational basis states.  

Our QCBM circuits are PQCs constructed by alternating layers of rotation gates ($\lbrace g_i \rbrace$) with layers of two qubit entangling operations ($\lbrace \ell_i \rbrace$). The rotation layers are composed of either parameterized Y-axis rotations ($\mathrm{R_Y}(\theta)$) or with the decomposition ($\mathrm{R_Z}(\alpha) \mathrm{R_X}(\beta) \mathrm{R_Z}(\gamma)$).  We will refer to the QCBM parameterized with $\mathrm{R_Y}$ gates as \textit{single gate parameterized QCBM} (SGP-QCBM), and will refer to the QCBM parameterized with the $3$ gate decomposition as \textit{arbitrary gate rotation QCBM} (AGP-QCBM).  The entangling layers are constructed either by $n$ CNOT gates between nearest neighbor qubits, with periodic closure (PC) or using alternating sets of $n/2$ CNOTs, such that the rotation and entangling layer form unitary two-designs (2D) \cite{cerezo2021cost}.

\subsection{DDCL training}
\label{sec:ddcl_training}
DDCL is a variational quantum algorithm that was introduced in Reference \cite{benedetti2018generative} and has been used to train QCBM models \cite{liu2018differentiable,hamilton2019generative,leyton2019robust,Benedetti2019ddqcl, Benedetti2019PQC}.  It is a hybrid quantum-classical workflow where the parameters of a QCBM ansatz (the rotational angles of the single qubit gates) are trained using gradient-free or gradient-based optimization methods.  

The loss function used in DDCL measures the similarity between two distributions:  the fixed target distribution $\mathrm{P_{target}}$ and the distribution sampled from the QCBM $Q(\THETAVEC{})$.  For gradient-based training the circuit gradient is evaluated using the parameter shift rule \cite{schuld2019evaluating}.

\begin{figure}
    \centering
    \includegraphics[width=\linewidth]{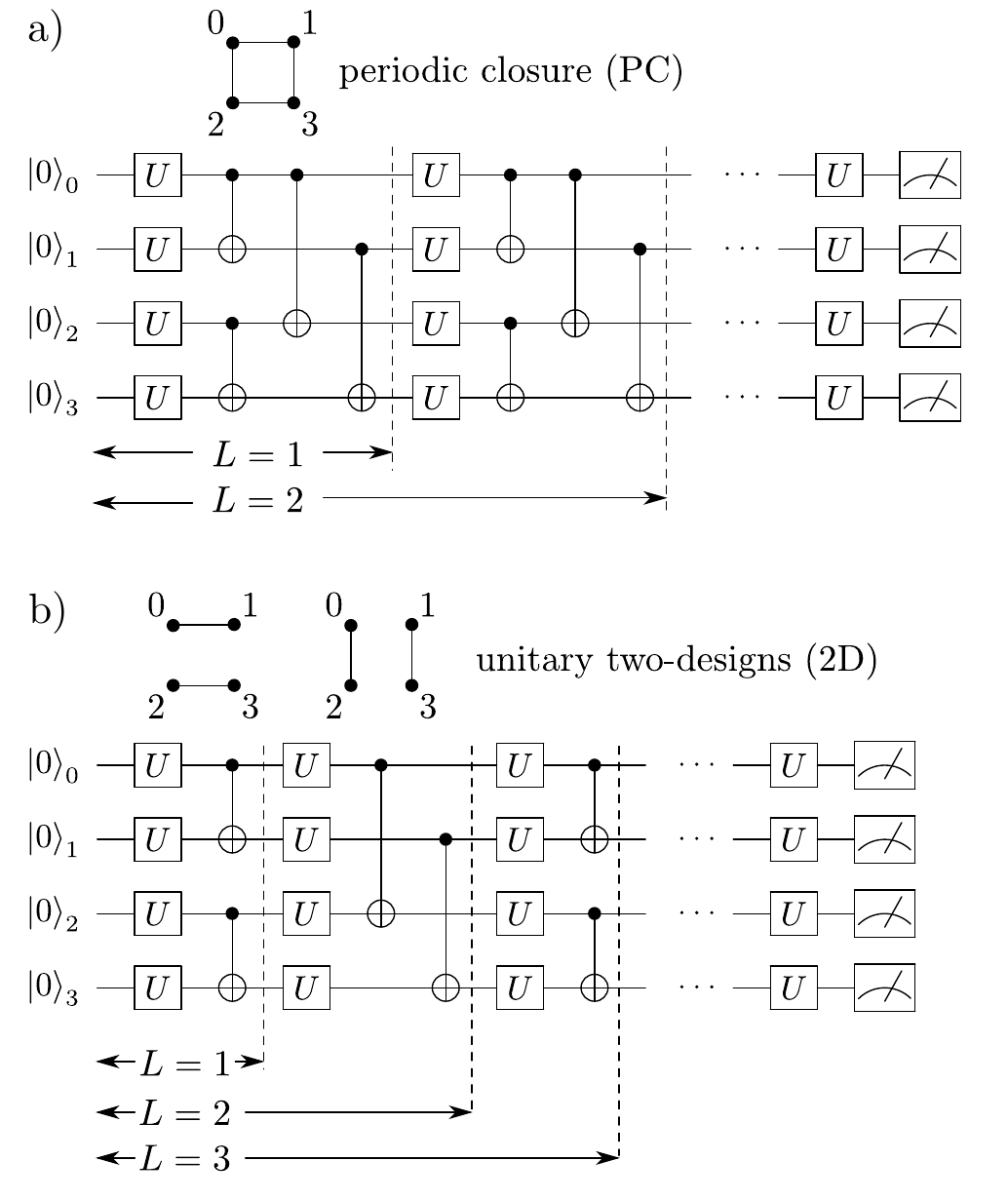}
    \caption{{\it PQC layouts for DDCL training:} In this figure we present the 4-qubit circuit layout for the QCBM parameterized with $U$. In this work, we consider two cases, $U = R_{Y}(\theta)$ (SGP-QCBM) and, $U = R_{Z}(\alpha)R_X(\beta) R_Z(\gamma)$ (AGP-QCBM). In a) and b) we show the PC and 2D designs for the entangling layers, respectively. Additionally, we show how the depth of the circuit $L$ represents the number of layers composed by rotations and CNOTs gate. We always consider a rotation layer before measurement regardless the $L$-level, this last layer allows us to consider the trivial case $L=0$. The pattern showed in these layouts can be extended for an arbitrary number of qubits ($>1$).}
    \label{fig:circuits}
\end{figure}
\subsection{Mean shift clustering}
\label{sec:clustering}
Each QCBM, with parameterization $\lbrace g_i\rbrace$, $\ell$ layers, and entangling layout $L_i$ is trained multiple times. We extract a set of representative minima from these multiple training data sets using mean shift clustering. In our tests we estimate the bandwidth of our parameter sets using radial basis functions with quantile $0.1$.  The set of $50$ $N$-dimensional vectors are reduced to a smaller subset $\lbrace m \rbrace$.  The set $\lbrace m \rbrace$ is further reduced to $\lbrace m^{\prime} \rbrace$ by applying down selection to limit to those centers which correspond to a loss value which is within a $95\%$ confidence interval of the mean final loss (defined using a t-test).

\subsection{Landscape visualization}
\label{sec:visualization}
Our visualization of the QCBM loss landscape relies on contour maps. Apart from the $N=2$, $d=0$, $\lbrace g_i \rbrace = \mathrm{R_Y}$ model, all our QCBMs contain $|\THETAVEC{}|>2$ parameters. Our contour plots are projections of this high-dimensional parameter spaces onto a 2-dimensional plane.  This plane is constructed using three parameter vectors $\THETAVEC{a},\THETAVEC{b},\THETAVEC{c}$ and the standard Gram Schmidt construction of two orthogonal unit vectors.  The first unit vector is directed from $w_1 \sim \THETAVEC{a}\to \THETAVEC{b}$ and the second unit vector ($w_2$) is constructed to be orthogonal to $w_1$ and chosen to intersect with \THETAVEC{c}.  For the edge case of $\THETAVEC{a}=\THETAVEC{b}=\THETAVEC{c}$ this construction method would fail.  However, the noisy training (noise induced by shot size) of QCBMs makes it rare that multiple training runs converge to the exact same parameter set.  There is the possibility that during the mean shift clustering and down selection (described in Section \ref{sec:clustering}) the algorithms could return cluster centers that are close in parameter space.

With the two orthogonal vectors defined, a grid of points is generated using NumPy's \texttt{meshgrid} function.  Along each dimension, a set of interpolated parameter values are defined. For example, translation along the $w_1$ axis is done by interpolating between \THETAVEC{a} and \THETAVEC{b}, similar to the interpolation methods described in \cite{goodfellow2014qualitatively,li2017visualizing}. Thus our contour plots are generated by projecting the high dimensional loss landscape onto a discrete set of equally spaced points in the $(w_1,w_2)$ plane. Each loss value is generated by updating all the parameters in \THETAVEC{} according to the directions defined by $w_1$ and $w_2$.

In \cite{goodfellow2014qualitatively}, the interpolated parameter sets are defined between two points (\THETAVEC{A}, \THETAVEC{B}) as
\begin{equation}
    \THETAVEC{\alpha} = \alpha\THETAVEC{A} + (1-\alpha)\THETAVEC{B},
\end{equation}
where $\alpha$ is constrained to the interval $[0,1]$. For our landscape visualization we expand the range for the parameter $\alpha$.  

\subsection{Nudged elastic band analysis}
\label{sec:neb_methods}
When analyzing the loss landscape of QCBM models, we are interested in quantifying the connectivity between regions of low loss.  Our over-parameterized QCBM circuits, and the fact that the training only uses the sampled distribution in evaluating the loss function, lead to a high degree of redundancy in our loss landscape.  Multiple parameter sets can correspond to low loss values either because they prepare quantum states with equivalent distributions when sampled in the computational basis or because over-parameterization can lead to redundant parameters. 

The nudged elastic band (NEB) algorithm \cite{henkelman2000neb} searches for low loss paths in high dimensional spaces.  The initial path connecting two known minima is a straight-line of interpolated values, discretized  into ($t$) linear segments.  The positions of the two known minima remain fixed, but the endpoints of the linear segments are updated using stochastic gradient descent.  As a result, the initial straight line path connecting two minima is stretched and deformed into a piece-wise linear path.  While our landscape visualization and NEB both rely on linearly interpolated paths, we will denote the discrete points along a fixed straight line path using $\alpha, \beta$, and we will denote the pivot points along a NEB piece-wise linear curve using $t$.
\section{QCBM Training Results}
\label{sec:QCBM_training}

We use multiple trainings to explore the loss landscape of SGP-QCBM and AGP-QCBM circuits (see Section \ref{sec:qcbm_construction}).  The maximum number of entangling layers is $4$ and we include the case of $0$ entangling layers as a control case.  The target distribution is derived from the n-qubit GHZ state and requires all qubits in the register to be fully entangled.  By including the case of circuits with $0$ entangling layers we can probe characteristics of the landscape of a circuit which cannot possibly fit the target distribution.  
\begin{figure*}[htbp]
\centering
\includegraphics[width=\textwidth]{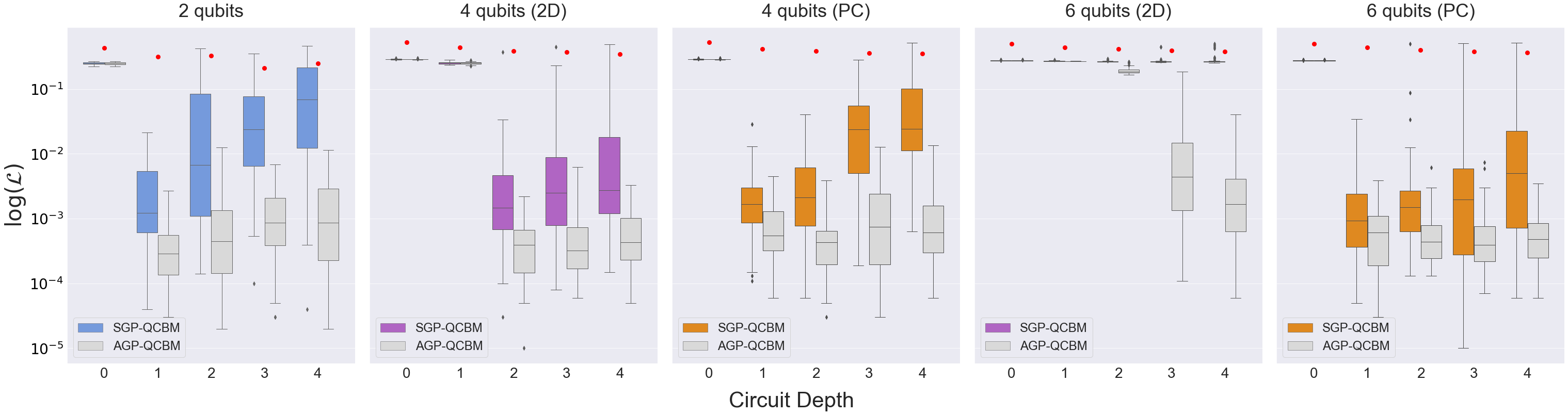}
\caption{Distribution of loss $\log{\mathcal{L}}$ values for all QCBMs trained in this study.  Values taken from the final step of optimization with Adam. The red markers (circles) denote the mean initial loss value computed over both parameterizations.}
\label{fig:log_loss_plot}
\end{figure*}

Landscape exploration is done through brute force sampling:  each circuit, parameterized by a gate set $\lbrace g_i \rbrace$, with $n$ qubits and $L$ entangling layers is trained $100$ individual times.  For $25$ out of these $50$ training runs we fix the initialization to $\THETAVEC{} = [0, 0, \dots , 0]$ \cite{grant2019initialization}.  The remaining $50$ training sets are done with parameters drawn uniformly at random from $[0, 2\pi)$. 

Each QCBM was trained with $50$ steps of Adam, an adaptive stochastic gradient descent method \cite{kingma2014adam}.  The training minimized the maximum mean discrepancy loss function \cite{liu2018differentiable,hamilton2019generative} and the loss was recorded at each optimization step.  The training was done in Qiskit \cite{Qiskit} using the noiseless qubit simulator.  During training each distribution was measured using $n_s = 2048$ shots.  In Fig.~\ref{fig:log_loss_plot} we show the distribution of final loss values for each QCBM. 

\section{Landscape Characterization}
\label{sec:landscape_characterization}
In this section we characterize the loss landscape of QCBM with a random exploration of the landscape, and connectivity analysis. From the training results plotted in Fig. \ref{fig:log_loss_plot}, we use our landscape analysis to understand how the three design choices: the parameterization, the circuit depth and the entangling design affect the performance of the final QCBM model.

We trained multiple QCBM circuits to fit the classical distribution found by sampling a $n$-qubit GHZ state.  Since we are using the distribution over computational basis states, this induces a degree of degeneracy in our landscape, by discarding the phase information in our final prepared states we have multiple states that can result in the same distribution. When appropriate, we will indicate whether a minimum is closer to the state $|\mathrm{GHZ+}\rangle = 1/\sqrt{2}(|00\dots0\rangle + |11\dots1\rangle$ or $|\mathrm{GHZ-}\rangle = 1/\sqrt{2}(|00\dots0\rangle - |11\dots1\rangle$.  Additionally, as the number of circuit parameters increases, the number of unique parameter sets that can prepare the target distribution increases.  Finally, due to the periodic nature of the parameterized rotation gates, there is rotational symmetry in the parameter sets. Overall this leads to a landscape that has many degenerate minima.  

\subsection{Landscape exploration}
\label{sec:exploration}
We begin with a simple examination of all pairwise combinations of minima found in the landscape using straight-line interpolated paths between pairs of minima.  In Tables \ref{tab:landscape_minima_2Q}, \ref{tab:landscape_minima_4Q} and \ref{tab:landscape_minima_6Q} we report the lowest barriers found between pairs of minima.  The barrier height is defined as the maximum loss value found along a straight-line path connecting two minima \THETAVEC{A}, \THETAVEC{B}.  We report the barrier heights with respect to the depth of the minima:
$\alpha = (\max{\mathcal{L}_{A\to B}}-(\min{\mathcal{L}_{A},\mathcal{L}_{B}}))/\sqrt{n_s})$. The number of shots is fixed at $n_s=8192$. 

\begin{table}[!t]
\renewcommand{\arraystretch}{1.3}
\caption{Number of unique minima ($m$) found in the $2$ qubit QCBM landscape using $50$ training runs according to mean shift clustering, the final set of unique minima after down selection ($m^{\prime}$), and the lowest barrier height ($\alpha$).}
\label{tab:landscape_minima_2Q}
\centering
\begin{tabular}{|c|c|c|c|c|}
 \hline
  QCBM & L & m & $m^{\prime}$ & $\alpha$\\
 \hline
 \multirow{5}{*}{SGP} & 0 & 7 & 6  & 0.32\\
                    & 1 & 20  & 18 & 0.46\\                      
                    & 2 & 27  & 15 & 0.10\\
                    & 3 & 16  & 9 & 0.15\\                      
                    & 4 & 17  & 6 & 0.18\\
\hline
                     \multirow{5}{*}{AGP} & 0 & 13 & 10 & 0.31  \\    
                    & 1 & 18  & 10 & 6.20\\
                    & 2 & 25  & 21 & 2.41\\             
                    & 3 & 25  & 20 & 2.40\\
                    & 4 & 26  & 20 & 1.38\\                      
\hline
\end{tabular}
\end{table}
\begin{table}[!t]
\renewcommand{\arraystretch}{1.3}
\caption{Number of unique minima ($m$) found in the $4$ qubit QCBM landscape using $50$ training runs according to mean shift clustering, the final set of unique minima after down selection ($m^{\prime}$), and the lowest barrier height ($\alpha$).}
\label{tab:landscape_minima_4Q}
\centering
\begin{tabular}{|c|c|c|c|c|c|}
 \hline
Layout &  QCBM & L & m & $m^{\prime}$ & $\alpha$ \\
\hline
\multirow{2}{*}{-} & SGP & 0 & 11  & 5 & 0.03\\ 
 & AGP & 0 & 15  & 8 & $3.0 \times 10^{-5}$\\ 
\hline
 \multirow{10}{*}{PC}  & \multirow{5}{*}{SGP} & 1 & 23 &     15 & 0.43 \\
                    && 2 &  26 &     18 & 0.78\\             
                    && 3 & 27 &     19 & 0.13\\
                    && 4 & 27 &     19 & 0.11\\ 
                    \cline{2-6}
                    & \multirow{5}{*}{AGP} & 1 & 19 &     13 & 5.55  \\             
                    && 2 & 23 &     19 & 6.81 \\             
                    && 3 & 26 &     19 & 4.66\\
                    && 4 & 26 &     21 & 3.44\\
\hline
\multirow{10}{*}{2D}  & \multirow{5}{*}{SGP} & 1 &  17 &      4 & $4.23 \times 10^{-3}$\\             
                    && 2 &  26 &     23 & 0.61\\
                    && 3 &  27 &     22 & 0.25\\ 
                    && 4 &  26 &     17 & 0.11\\
                    \cline{2-6}
                    & \multirow{5}{*}{AGP} & 1 & 16 &      9 & $6.08 \times 10^{-3}$\\         
                    && 2 & 26 &     19 & 5.78\\             
                    && 3 & 25 &     20 & 7.53\\
                    && 4 & 26 &     17 & 3.30\\     
\hline
\end{tabular}
\end{table}
\begin{table}[!t]
\renewcommand{\arraystretch}{1.3}
\caption{Number of unique minima ($m$) found in the $6$ qubit QCBM landscape using $50$ training runs according to mean shift clustering, the final set of unique minima after down selection ($m^{\prime}$), and the lowest barrier height ($\alpha$).}
\label{tab:landscape_minima_6Q}
\centering
\begin{tabular}{|c|c|c|c|c|c|}
 \hline
Layout &  QCBM & L & m & $m^{\prime}$ & $\alpha$\\
\hline
\multirow{2}{*}{-} & SGP & 0 & 15  & 9 &$9.48 \times 10^{-3}$\\ 
 & AGP & 0 & 14  & 8 &$7.0 \times 10^{-5}$\\ 
\hline
\multirow{10}{*}{PC}  & \multirow{5}{*}{SGP} & 1 & 26 & 20 & 1.80 \\
                    && 2 & 26 & 23 & 0.57   \\             
                    && 3 & 26 & 18& 0.54    \\
                    && 4 & 26 & 17 &0.06  \\ 
                    \cline{2-6}
                    & \multirow{5}{*}{AGP} & 1 & 24 & 19 & 6.19    \\             
                    && 2 & 26 & 24 & 8.44  \\             
                    && 3 & 26 & 22& 7.18  \\
                    && 4 & 26 & 16 & 8.11  \\                      
\hline
 \multirow{10}{*}{2D}  & \multirow{5}{*}{SGP}& 1 &  23 & 11 & $8.88 \times 10^{-3}$ \\             
                    && 2 &  24 & 14 & $8.75 \times 10^{-3}$ \\
                    && 3 &  26 & 22 & $9.25 \times 10^{-3}$ \\ 
                    && 4 &  26 & 18 & $9.51 \times 10^{-3}$  \\
                    \cline{2-6}
                    & \multirow{5}{*}{AGP} & 1 & 26 & 19 &$4.96 \times 10^{-3}$ \\         
                    && 2 & 26 & 25 & 0.02 \\             
                    && 3 & 26 & 22 & 0.74 \\
                    && 4 & 26 & 25& 4.18 \\                       
\hline
\end{tabular}
\end{table}

The barrier heights reported in Tables \ref{tab:landscape_minima_2Q}, \ref{tab:landscape_minima_4Q} and \ref{tab:landscape_minima_6Q} are encountered when a straight-line of interpolated values is defined between two minima.  This is not an accurate portrayal of how the loss landscape is traversed during training.  Additionally, fully characterizing the local curvature in the loss landscape requires the computation of the eigenspectra associated with the loss function Hessian \cite{sagun2016eigenvalues,huembeli2021characterizing,mari2021gradient}. However, these tests can be instructive in terms of identifying large-scale features in the loss landscape.  

\subsection{Parameterization and QCBM performance}
\label{sec:parameterization}
The values reported in Tables \ref{tab:landscape_minima_2Q}, \ref{tab:landscape_minima_4Q} and \ref{tab:landscape_minima_6Q} indicate that large barriers in the QCBM landscape exist, and that large barriers are common for the AGP-QCBM.  Yet in the log loss plot (Fig. \ref{fig:log_loss_plot}) we observe that the AGP-QCBM reaches final loss values that are orders of magnitude lower than those for the SGP-QCBM. We use NEB \cite{henkelman2000neb} to investigate if low loss paths can exist around these barriers.   

\begin{figure}[htbp]
\centering
\includegraphics[width=\columnwidth]{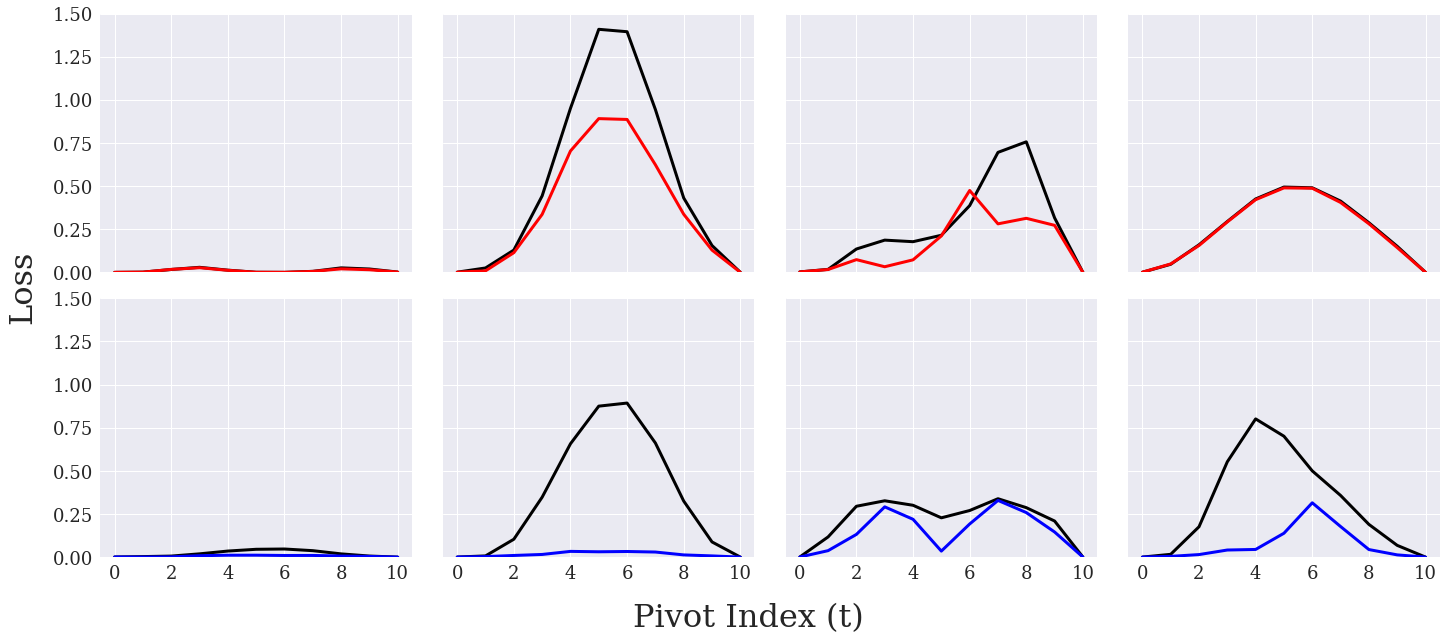}
\caption{Comparison of NEB curves with 2 qubit QCBM ($L = 1$) from best to worst performance.  (Top row) curves found for SGP-QCBM: initial curves (black), final curves (red) (Bottom row) curves found for AGP-QCBM: initial curves (black), final curves (blue).}
\label{fig:NEB_2Q_sample_D1}
\end{figure}

We start with the 2 qubit QCBM, which has only 1 possible entangling layer design.  For our NEB analysis, we randomly sample  pairwise combinations of down selected minima $\lbrace m^{\prime} \rbrace$. In Fig. \ref{fig:NEB_2Q_sample_D1} we show a representative set of plots from the 2 qubit QCBM landscape with $L=1$. We perform an abbreviated NEB search using 20 steps of gradient-based updates.  From this abbreviated search and   random sampling, we were able to observe several characteristic NEB behaviors around barriers.  We consider 4 characteristic behaviors:  the barrier between two endpoints could be low, and the landscape is relatively flat between two minima; the barrier could be high and NEB is able find a path around it;  the barrier could be high and NEB is only able to find a wide saddle between the endpoints; NEB fails to make any improvements or reduction in the loss.  

\begin{figure}[htbp]
\centering
\includegraphics[width=\columnwidth]{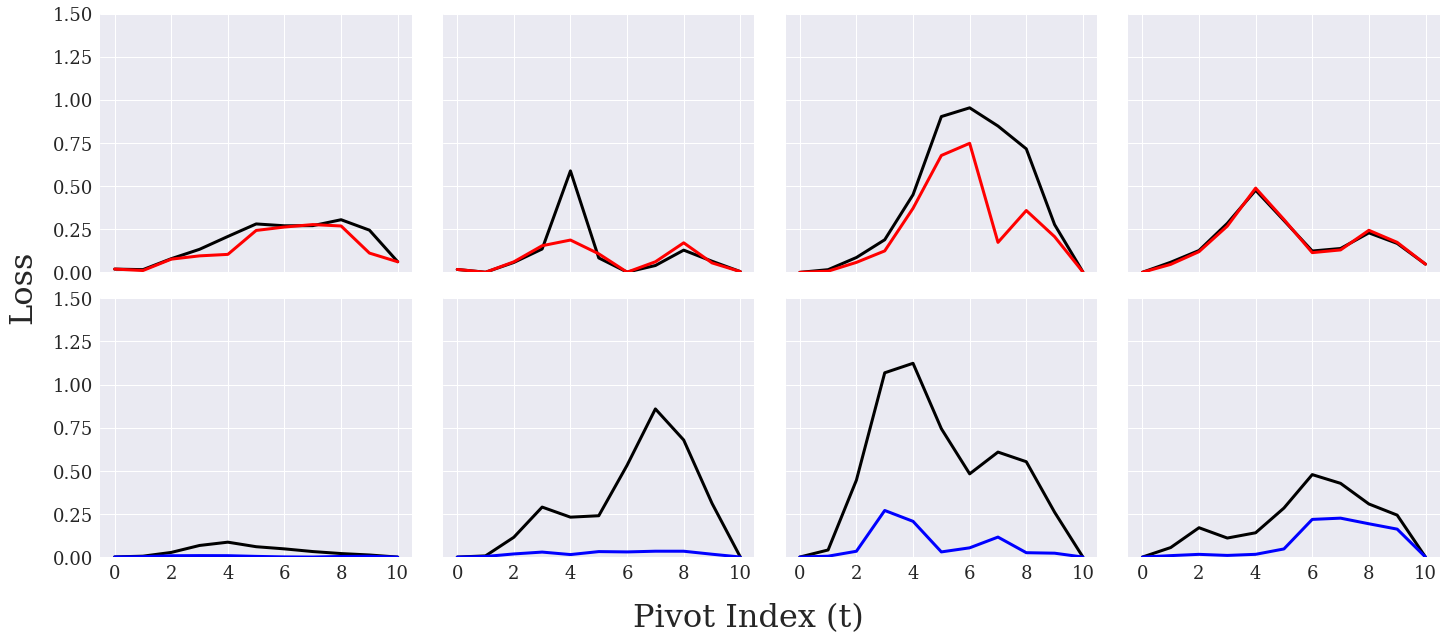}
\caption{Comparison of NEB curves with 2 qubit QCBM ($L = 2$) from best to worst performance.  (Top row) curves found for SGP-QCBM: initial curves (black), final curves (red) (Bottom row) curves found for AGP-QCBM: initial curves (black), final curves (blue).}
\label{fig:NEB_2Q_sample_D2}
\end{figure}
\begin{figure}[htbp]
\centering
\includegraphics[width=\columnwidth]{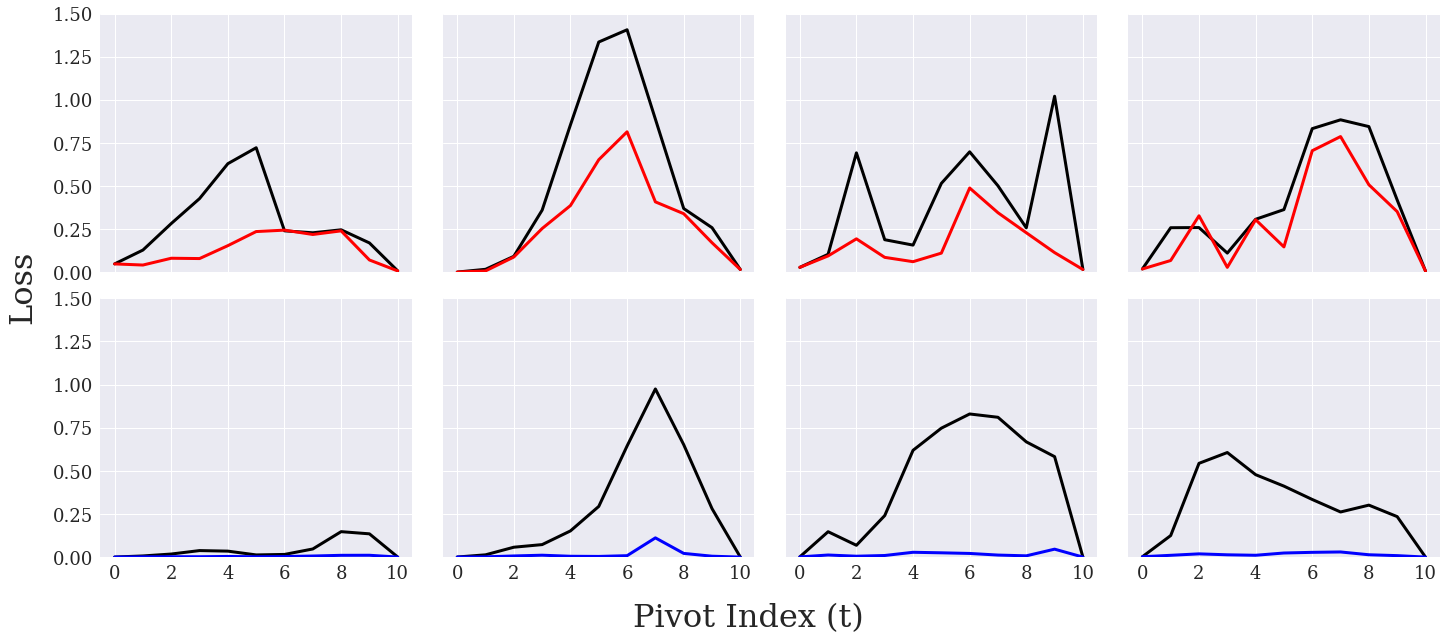}
\caption{Comparison of NEB curves with 2 qubit QCBM ($L = 3$) from best to worst performance.  (Top row) curves found for SGP-QCBM: initial curves (black), final curves (red) (Bottom row) curves found for AGP-QCBM: initial curves (black), final curves (blue).}
\label{fig:NEB_2Q_sample_D3}
\end{figure}
\begin{figure}[htbp]
\centering
\includegraphics[width=\columnwidth]{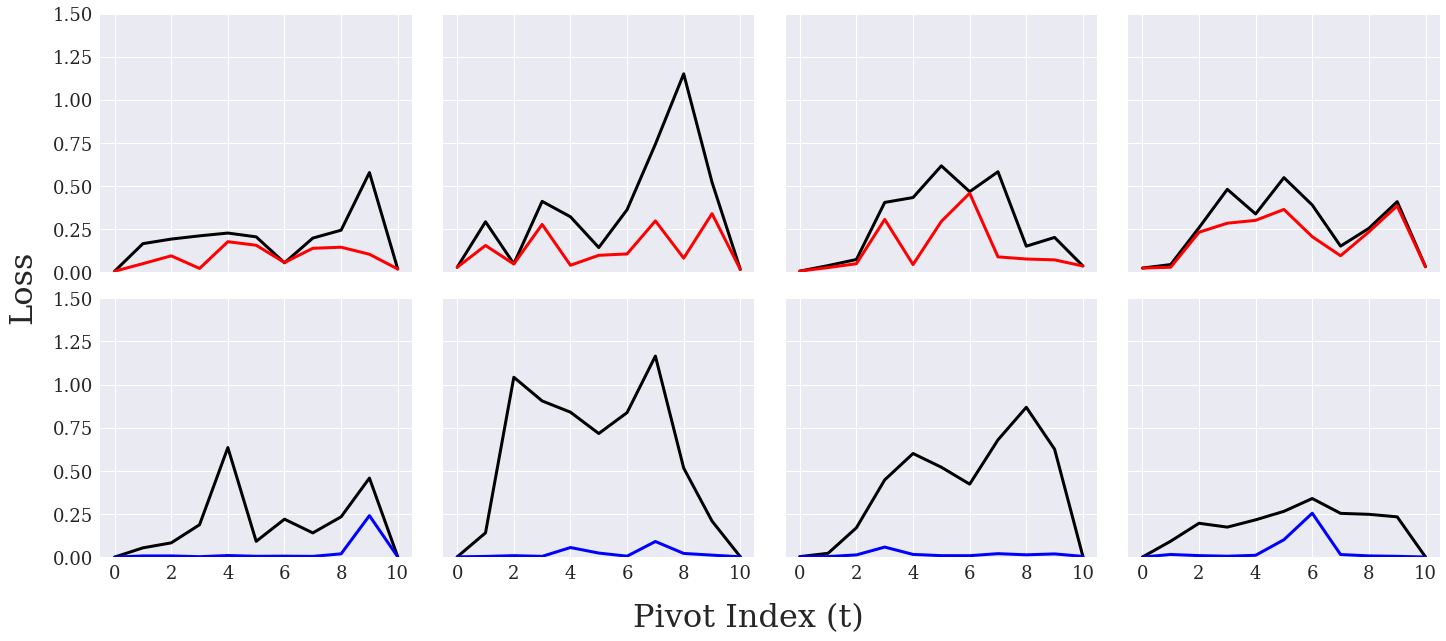}
\caption{Comparison of NEB curves with 2 qubit QCBM ($L = 4$) from best to worst performance.  (Top row) curves found for SGP-QCBM: initial curves (black), final curves (red) (Bottom row) curves found for AGP-QCBM: initial curves (black), final curves (blue).}
\label{fig:NEB_2Q_sample_D4}
\end{figure}

 Increasing the depth of the circuits benefits the AGP-QCBM, in Figs. \ref{fig:NEB_2Q_sample_D2}, \ref{fig:NEB_2Q_sample_D3} and \ref{fig:NEB_2Q_sample_D4} we show a representative sample of the NEB behaviour for QCBMs with $L=2$, $L=3$ and $L=4$, respectively. With a random sampling of the minima in the SGP-QCBM it was very easy to find examples of the NEB algorithm failing to find low loss paths, or adequately navigating around barriers.  On the other hand, in the AGP-QCBM landscape it was very easy to find examples of NEB easily navigating around barriers. 

\subsection{Entangling layer design}
\label{sec:entanglement}
\begin{figure}[htbp]
\centering
\includegraphics[width=\columnwidth]{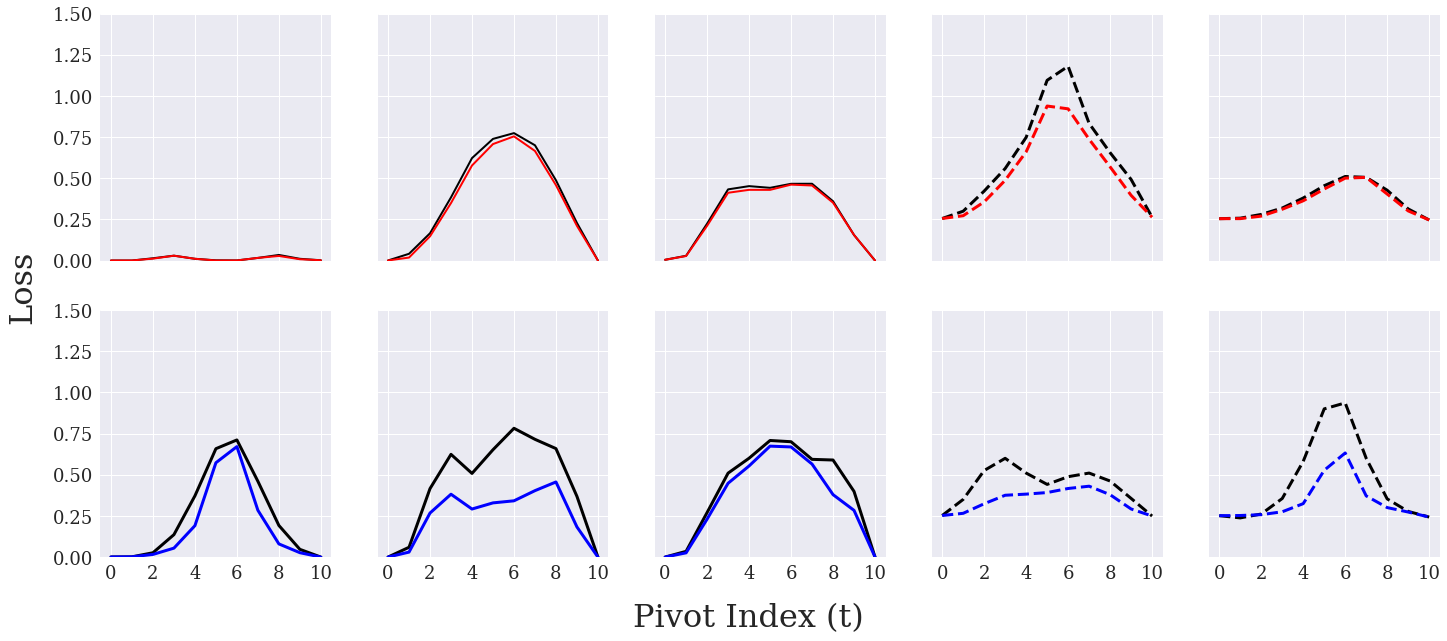}
\caption{Comparison of NEB curves with 4 qubit QCBM ($L = 1$).  (Top row) Curves found for SGP-QCBM (PC): initial curves (black, solid), final curves (red, solid). Curves found for SGP-QCBM (2D): initial curves (black, dashed), final curves (red, dashed). (Bottom row) Curves found for AGP-QCBM (PC): initial curves (black, solid), final curves (blue, solid). Curves found for AGP-QCBM (2D): initial curves (black, dashed), final curves (blue, dashed)}
\label{fig:D1_NEB_curves_Q4}
\end{figure}

As the width of the QCBM models is increased, the effects of circuit depth and entangling layer design are correlated.  All QCBM with $\mathrm{L}=0$ are incapable of preparing entangled states. Additionally, QCBMs constructed with the 2D layout cannot prepare the target distribution if $L = 1$. As the QCBM become larger, running a short NEB search (20 pivot updates) is not guaranteed to find a low loss path.  However, in Fig. \ref{fig:D1_NEB_curves_Q4} we can observe that with the PC layout, the SGP-QCBM again have difficulty modifying the original curve, while the AGP-QCBM can maneuver around barriers.

\begin{figure}[htbp]
\centering
\includegraphics[width=\columnwidth]{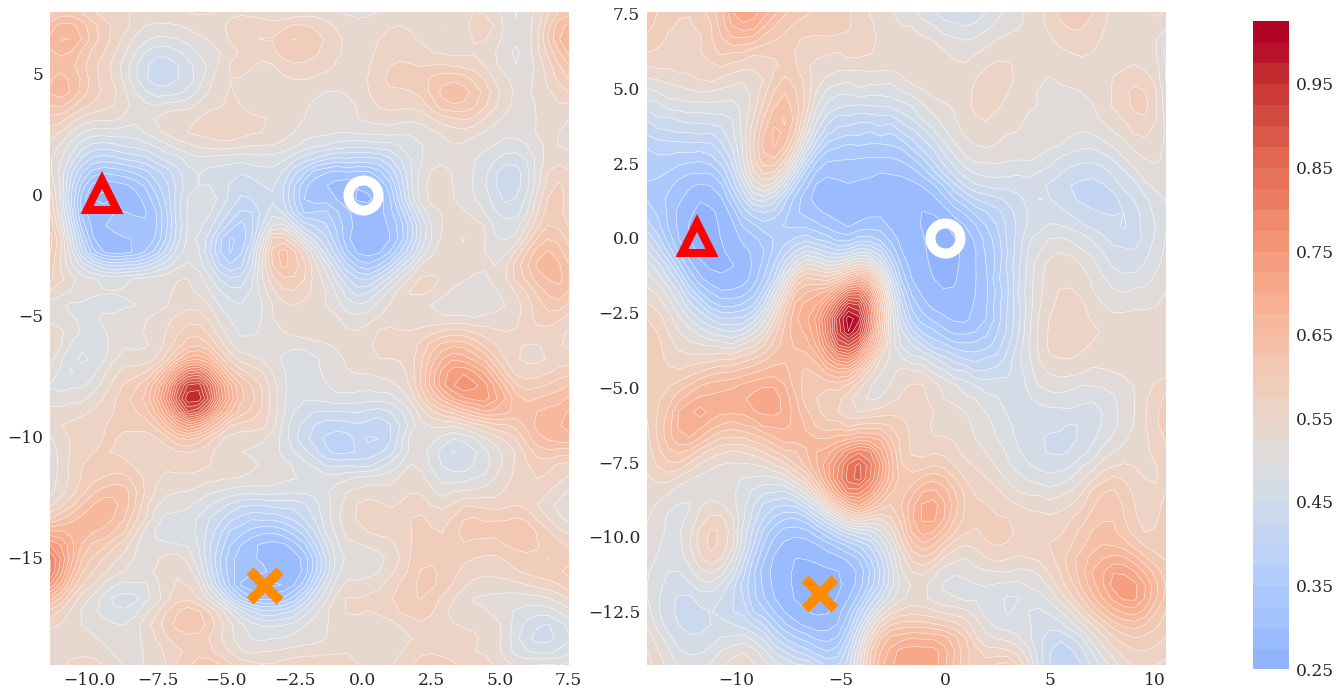}
\caption{Contour plots for 6 qubit QCBM with $L = 1$ and the 2D entangling layout.  (Left) SGP-QCBM, (Right) AGP-QCBM.}
\label{fig:D1_VC_contours_Q6}
\end{figure}

\begin{figure}[htbp]
\centering
\includegraphics[width=\columnwidth]{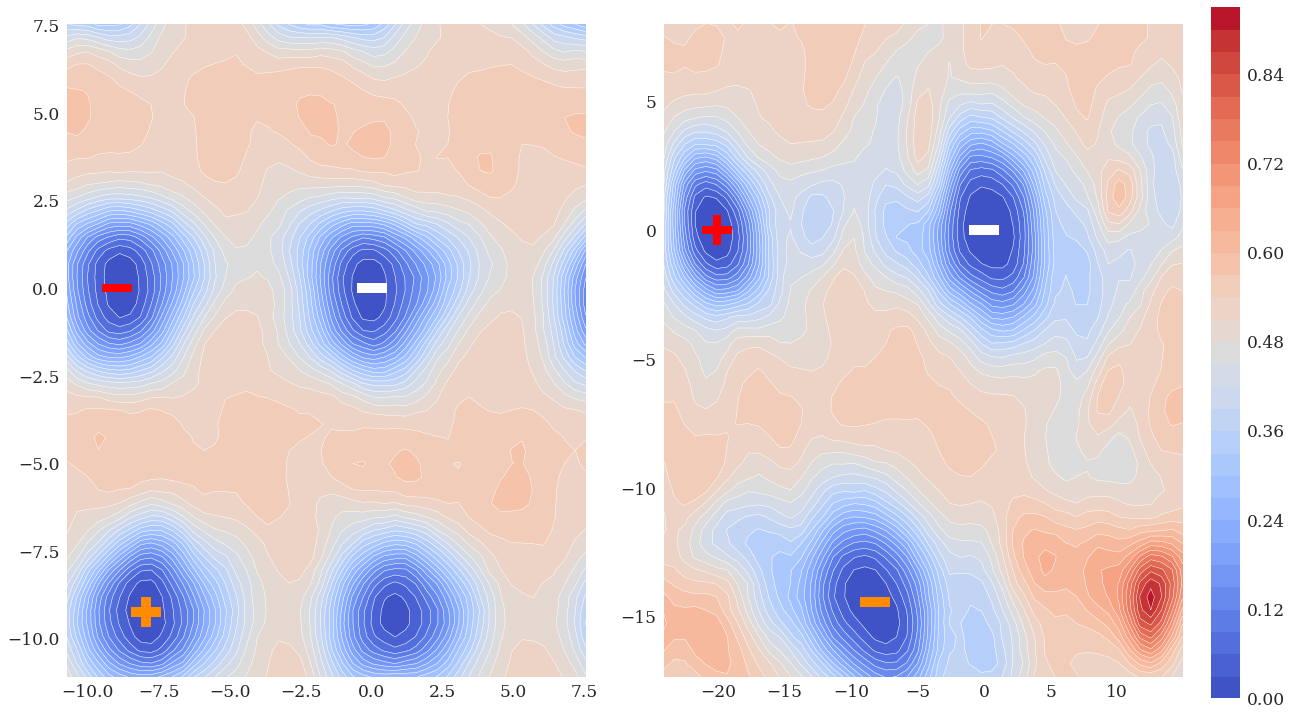}
\caption{Contour plots for 6 qubit QCBM with $L = 1$ and the PC entangling layout.  (Left) SGP-QCBM , (Right) AGP-QCBM. Marker shape denotes closest GHZ state ($|\mathrm{GHZ}+\rangle, |\mathrm{GHZ}-\rangle$).}
\label{fig:D1_SC_contours_Q6}
\end{figure}

As a comparison, consider the mean final losses found for the $L=1$ SGP-QCBM and AGP-QCBM as plotted in Fig. \ref{fig:log_loss_plot}. With the 2D entangling layout a $L=1$ QCBM lacks the entangling capacity to prepare either GHZ state.  The PC entangling layout does not face this problem.  With 4- or 6- qubits, the $L=1$ AGP-QCBM built with the PC entangling layout reach far lower final losses. In Figs. \ref{fig:D1_VC_contours_Q6} and \ref{fig:D1_SC_contours_Q6}, we plot a set of contours for the 6 qubit, $L=1$ SGP-QCBM (12 trainable parameters) and AGP-QCBM (24 trainable parameters).  For models that cannot prepare the target distribution the training converges to shallow local minima, regardless of the parameterization (Fig. \ref{fig:D1_VC_contours_Q6}). With the PC entangling layer, the local minima are deeper for both parameterizations (Fig. \ref{fig:D1_SC_contours_Q6}).

\subsection{Ravines}
\label{sec:ravines}
We continue our landscape characterization with several observations of ravines in the QCBM loss landscape.  Ravines are regions of the landscape where multiple minima can be connected by straight-line paths, along which the loss shows a minimal increase. The values in Tables \ref{tab:landscape_minima_2Q}--\ref{tab:landscape_minima_6Q} and the representative NEB plots (Figs. \ref{fig:NEB_2Q_sample_D1}) show that relative low barrier heights can exist between minima.  We use these low relative barrier heights, or flat NEB curves, to identify QCBMs that may contain ravines.  

\begin{figure}[htbp]
\centering
\includegraphics[width=\columnwidth]{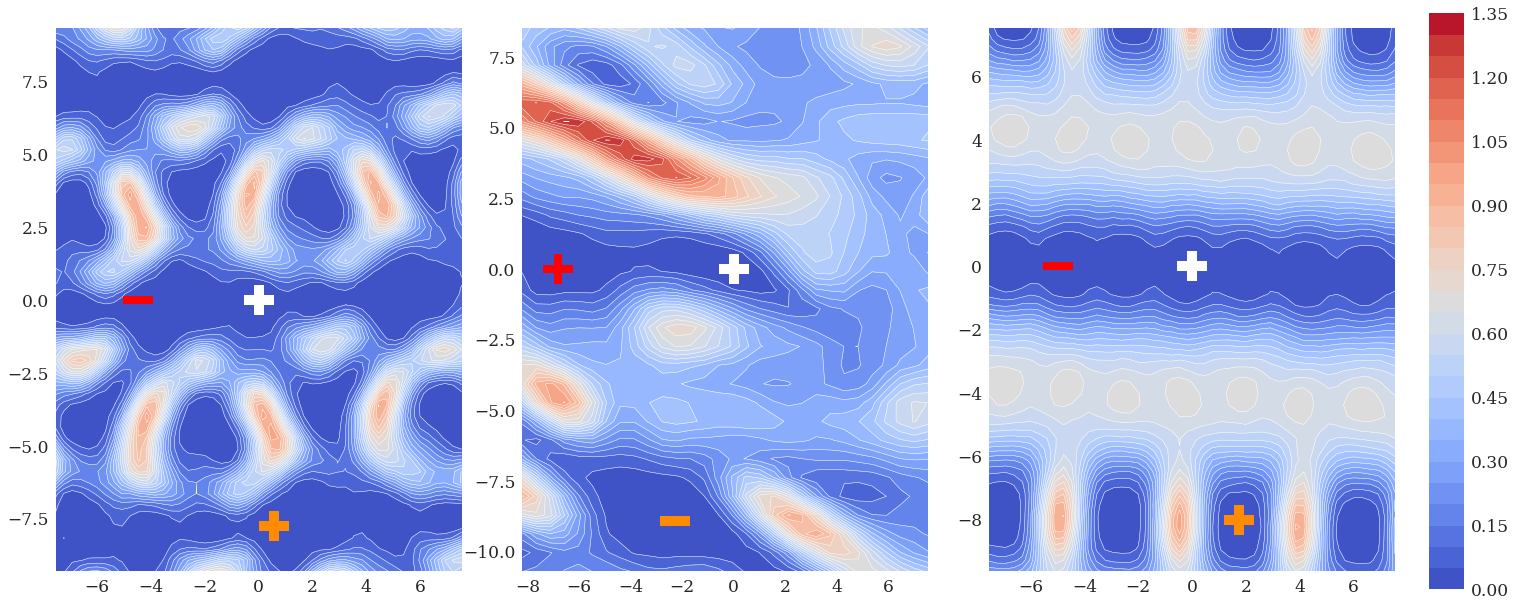}
\caption{ Contour plots for 3 ravines located by flat NEB curves  (Left)2 qubit SGP-QCBM with $L = 1$, (Center) 2 qubit AGP-QCBM with $L = 1$,  (Right) 4 qubit SGP-QCBM with $L = 1$. Marker shape denotes closest GHZ state ($|\mathrm{GHZ}+\rangle, |\mathrm{GHZ}-\rangle$).}
\label{fig:D1_contours_Q2}
\end{figure}

In Fig. \ref{fig:D1_contours_Q2} we plot a set of contours corresponding to potential ravines, as identified by flat NEB curves. Potential ravines were identified for the 2 qubit, $L=1$ QCBM (SGP and AGP), and the 4 qubit $L=1$ SGP-QCBM.  Two of the three points (\THETAVEC{A}, \THETAVEC{B}) plotted in the leftmost plot of Fig. \ref{fig:D1_contours_Q2} are the minima plotted in the leftmost column of Fig. \ref{fig:NEB_2Q_sample_D1}: these are examples of minima that are connected by nearly flat NEB curves.

\begin{figure}[htbp]
\centering
\includegraphics[width=\columnwidth]{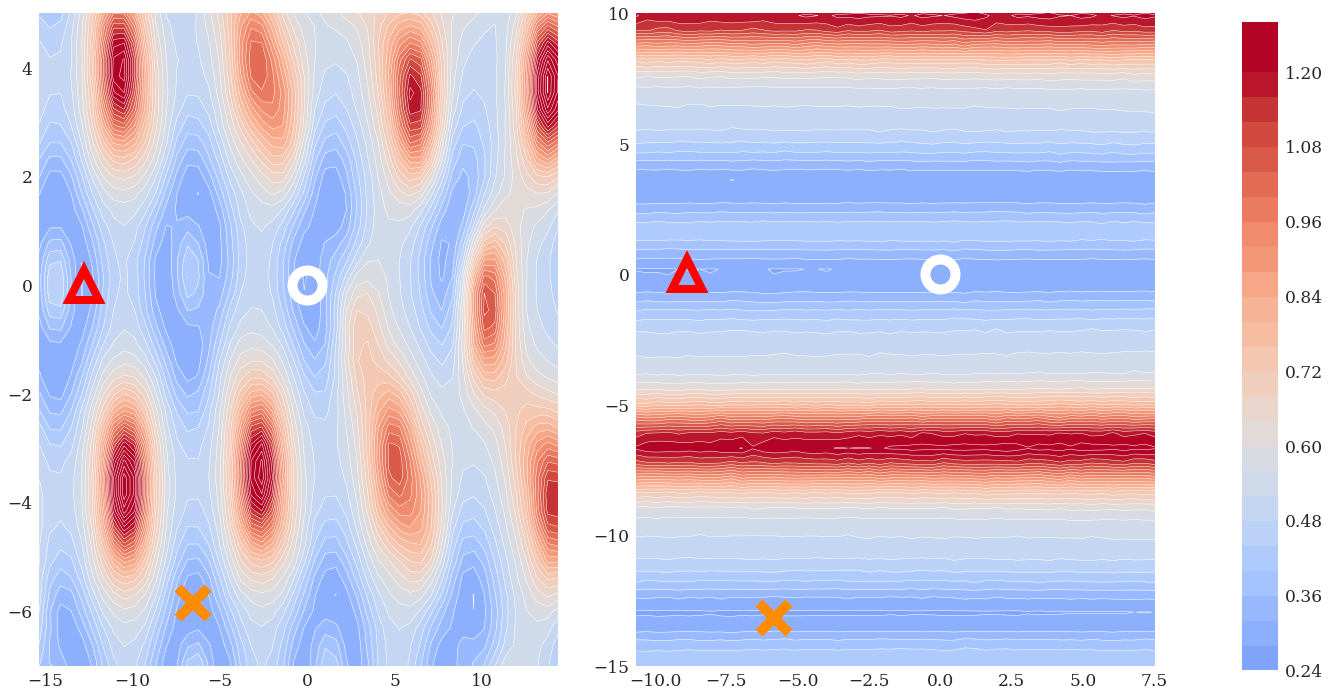}
\caption{Contour plots for 6 qubit QCBM with $L = 0$.  (Left) SGP-QCBM, (Right) AGP-QCBM. These models did not train sufficiently close to either GHZ state.}
\label{fig:D0_contours_Q6}
\end{figure}

In Fig. \ref{fig:D0_contours_Q6}, we plot a set of contours for the 6 qubit, $L=0$ QCBM.  The three points used to define the plane were chosen according to $\alpha$: \THETAVEC{A}, \THETAVEC{B} are the two minima that gave the value of $\alpha$ reported in Table \ref{tab:landscape_minima_6Q}; \THETAVEC{C} is a minimum that had the highest value of $\alpha$ with respect to \THETAVEC{A} or \THETAVEC{B}.  The low relative barrier heights for the AGP-QCBM corresponds to minima sitting in the same ravine, while the low relative barrier heights for the SGP-QCBM corresponds to periodic valleys.  The lowest value of the loss for either model does not go below $0.2$.

The importance and role of ravines in training remains an open question.  The existence of a ravine in a landscape does not guarantee that low losses can be achieved: we observe that ravines can be found in the trivial landscape of $L=0$ QCBM.  Rather, the occurrence of ravines may be closer linked to redundant parameters and dropout stability \cite{Hamilton2020visualization}. 

\subsection{Plateaus}
\begin{figure*}[htbp]
\centering
\includegraphics[width=\textwidth]{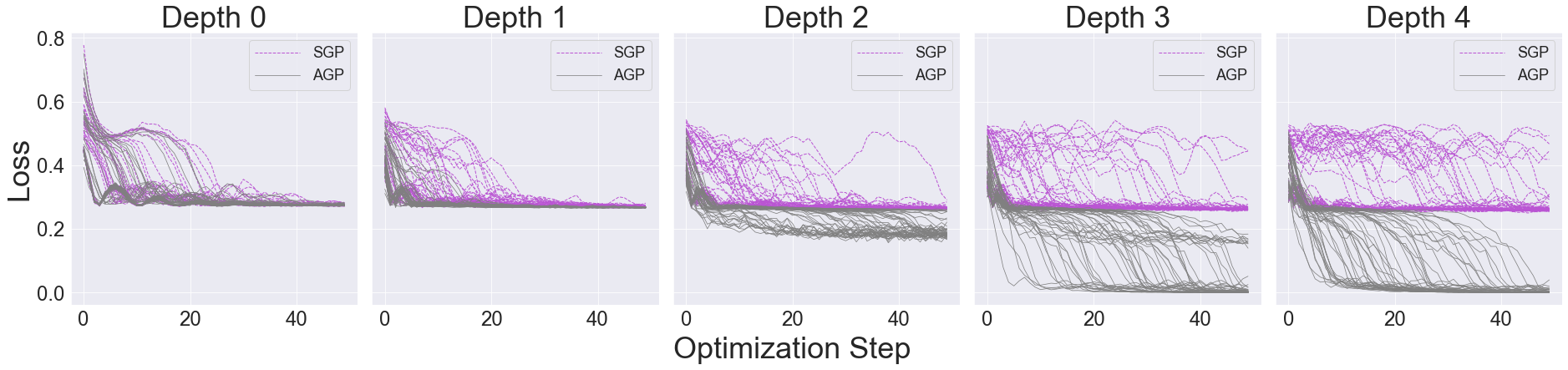}
\caption{Individual loss traces for 6 qubit QCBMs with 2D entangling layout.}
\label{fig:loss_trace_VC}
\end{figure*}

\begin{figure*}[htbp]
\centering
\includegraphics[width=\textwidth]{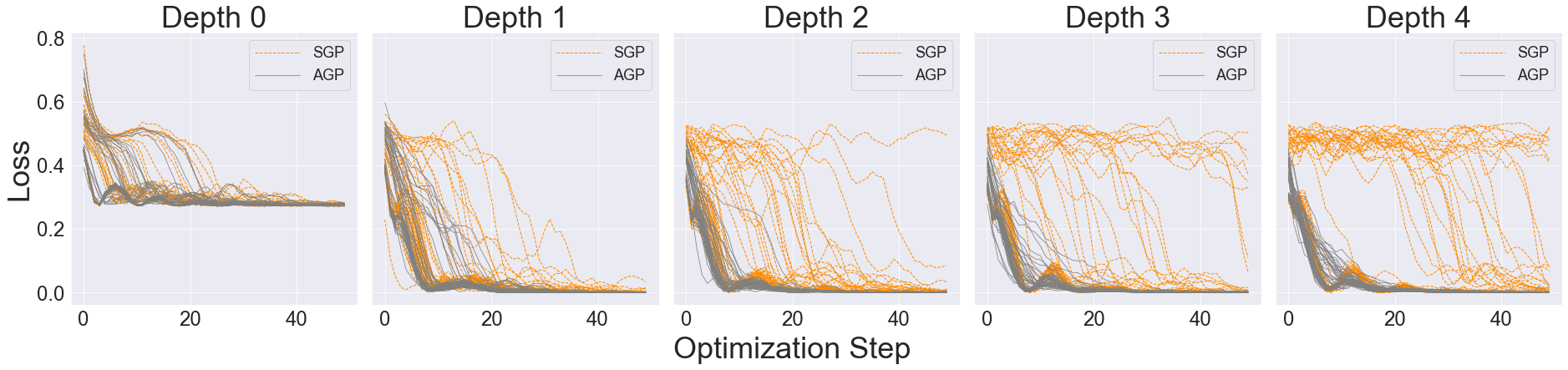}
\caption{Individual loss traces for 6 qubit QCBMs with PC entangling layout.}
\label{fig:loss_trace_SC}
\end{figure*}

Finally, we close this section on landscape characterization by investigating the poor training of larger QCBM models.  From the 2- and 4-qubit results, we observe a clear connection between parameterization and depth, and entangling layer design and depth. The 6- qubit results in Fig. \ref{fig:log_loss_plot} show a stronger correlation between the parameterization choice, entangler design and performance.  While we expect the $L=1$ QCBM models with 2D entangler design to train poorly, the 6-qubit results show that the combination of 2D entangler design and SGP-QCBM trains poorly at any depth.  To facilitate our analysis of these models, in Figs. \ref{fig:loss_trace_VC} and \ref{fig:loss_trace_SC} we plot the individual traces of the loss during training for the  6 qubit QCBM.  

Comparing the mean loss values plotted in Fig. \ref{fig:log_loss_plot} with the individual traces of the losses in Figures \ref{fig:loss_trace_VC} and \ref{fig:loss_trace_SC}, we note that the AGP-QCBM consistently reaches lower final loss values.  With the loss traces plotted on a linear scale, we can see in greater detail how the parameterization affects training.  Based on the loss trace plots we observe that training in the SGP-QCBM landscape with 2D entangling layout reaches a minimum loss on the order of 0.2 and fails to find a lower loss value.  While this floor is expected for any QCBM with $L \leq 1$ (due to lack of full entanglement) the minimum loss value persists even as the circuit depth is increased. 
Increasing the density of the entangling layer (PC layout) mitigates this behavior -- SGP circuits with $L\geq 1$ can reach lower loss values (Fig.~\ref{fig:loss_trace_SC}). 
However, as the length increases we observe a return of a plateau at a loss of approximately 0.5.  

On the other hand, in the AGP-QCBM with 2D entangling layout, the minimum loss on the order of 0.2 occurs for short circuits, but for $L>2$ the loss can be reduced further. From the individual loss traces we can begin to identify barren plateaus.  As the depth of the QCBM increases, the SGP-QCBM initialized with random values exhibit an initial plateau in the loss value: several optimization steps are required before the loss begins to decrease. Additionally there are several trainings which fail to minimize the loss. 

\begin{figure}[htbp]
\centering
\includegraphics[width=\columnwidth]{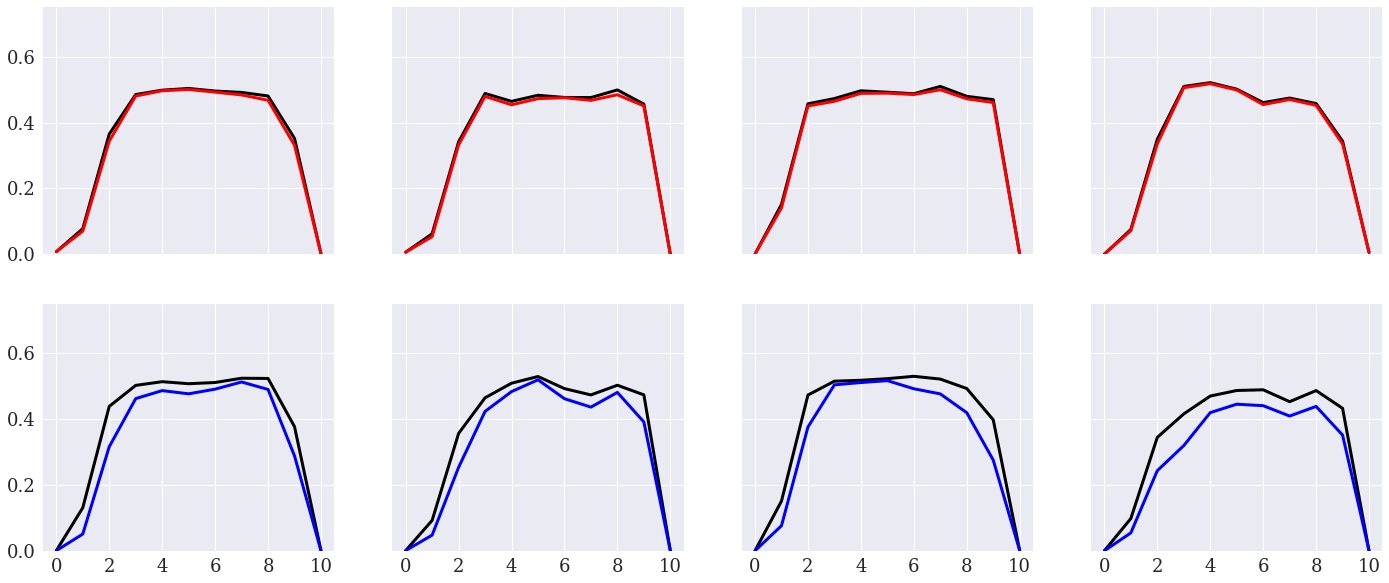}
\caption{Comparison of NEB curves with 6 qubit QCBM ($L = 3$), PC layout.  (Top row) curves found for SGP-QCBM: initial curves (black), final curves (red) (Bottom row) curves found for AGP-QCBM: initial curves (black), final curves (blue).}
\label{fig:NEB_6Q_sample_D3}
\end{figure}

In Section \ref{sec:parameterization} we showed that in the AGP-QCBM landscape, paths that circumvent large barriers are relatively easy to find.  In Section \ref{sec:entanglement}, we observed a minimum depth needed when using the 2D entangling layer design.  For the 6-qubit QCBMs we focus on the PC entanglement design.  In Fig. \ref{fig:NEB_6Q_sample_D3} we plot again a random selection of NEB curves for our trained QCBM models.  For the curves show, the minima correspond to trainings that have escaped from any plateau, but we can observe the qualitative changes in the landscape.  Compared to the NEB curves for 2- and 4-qubit QCBM, the landscape of the 6-qubit is becoming flatter - the barrier heights are much lower than what was seen in previous plots. Additionally, the barriers are wider then those seen in previous figures. Again, the SGP-QCBM shows minima changes in the loss along the NEB curve, whereas the AGP-QCBM shows small reductions in the barrier height.  As the QCBM landscape dimension increases, the NEB analysis may require more steps to update the pivot locations. 

\section{Conclusion}
\label{sec:conclusions}

In this work we have focused on several design choices that have been utilized as means to address several near-term challenges of working with quantum hardware:  reducing gate noise and reducing computational overhead.  The 2D entangling layout uses the fewest two-qubit gates in its design, and one would expect that when these circuits are executed on hardware they will incur lower amounts of two-qubit gate noise.  However, this entangling layer choice has a minimum circuit depth required to ensure that your model can sufficiently fit a particular target distribution.  The use of single parameterized quantum gates (SGP-QCBM) can reduce the computational overhead by reducing the dimensionality of the circuit gradient.  However, the slight savings in circuit executions is offset by an increased rigidity in the model.  Additionally, as more qubits become available on quantum hardware, even with improved coherence times and lower gate noise, the size of parameterized circuits that can be built and trained on hardware will grow. Future work will need to focus on how hardware noise affects the connectivity of low loss paths in the QCBM landscape. 

Several recent studies have quantified the ``expressibility'' and ``trainability'' of commonly used ans\"atze \cite{sim2019expressibility,holmes2021connecting}. We focused on the loss landscape to identify two features (large barriers and plateaus) that could pose significant challenges to QCBM training, and how connectivity in the landscape can mitigate these challenges. The abbreviated NEB search was able to find low loss paths around large barriers for AGP-QCBM. SGP-QCBM built with the 2D entangling layout had worst NEB performance.  Combing these design choices led to landscapes with minima that appear to be separated by wide, flat plateaus. Yet, the fact that NEB can find low loss paths between several minima in the QCBM landscape and the connections to QCBM design is promising: it indicates that circuit design may be influential in creating landscapes that are optimal for training.  

\section*{Acknowledgment}

This work was supported as part of the ASCR Testbed Pathfinder Program at Oak Ridge National Laboratory under FWP ERKJ332.  This work was supported as part of the ASCR Fundamental Algorithmic Research for Quantum Computing Program at Oak Ridge National Laboratory under FWP ERKJ354. This work was partially supported as part of the ASCR QCAT Program at Oak Ridge National Laboratory under FWP \#ERKJ347. EL was supported by the U.S. Department of Energy, Office of Science, Office of Workforce Development for Teachers and Scientists (WDTS) under the Science Undergraduate Laboratory Internship program. S.M. is funded in part by an NSF QISE-NET fellowship (DMR-1747426).




\end{document}